\documentclass[english]{article}
\usepackage[utf8]{inputenc}
\usepackage[T1]{fontenc}
\usepackage[latin9]{luainputenc}
\usepackage{babel}
\usepackage{natbib}
\usepackage{amsfonts,amsmath,amsthm,amssymb}
\usepackage{bbm}
\usepackage{bigints}
\usepackage{graphicx}
\usepackage{systeme}
\usepackage{float}
\usepackage[margin=1in]{geometry}
\newtheorem{theorem}{Theorem}
\newcommand{\isEquivTo}[1]{\underset{#1}{\sim}}

\usepackage{mathtools}
\DeclarePairedDelimiter\abs{\lvert}{\rvert}

\DeclarePairedDelimiter\floor{\lfloor}{\rfloor}

\usepackage[toc]{appendix}

\newtheorem{corollary}{Corollary}

\title{Price Impact in a Latent Order Book}
\author{Ismael Lemhadri
\footnote{lemhadri@stanford.edu. 
}
}
\date{Department of Statistics - Stanford University}

\begin{document}

\maketitle

\begin{abstract}
The latent order book of \cite{donier2015fully} is one of the most promising agent-based models for market impact. This work extends the minimal model by allowing agents to exhibit mean-reversion, a commonly observed pattern in real markets. 
This modification leads to new order book dynamics, which we explicitly study and analyze. Underlying our analysis is a mean-field assumption that views the order book through its \textit{average} density. We show how price impact develops in this new model, providing a flexible family of solutions that can potentially be calibrated to real data. While no closed-form solution is provided, we complement our theoretical investigation with extensive numerical results, including a simulation scheme for the entire order book.
\end{abstract}

\newpage
\tableofcontents

\newpage

\section{Introduction}\label{introduction}
In modern financial markets, the concept of \textit{liquidity risk} reflects the extra-cost incurred by a trader that is fundamentally due to the scarcity of supply. In the most extreme cases, this can make trading nearly impossible in the absence of counterparty. One striking example occurred in 2007 during the subprime mortgage crisis when products such as collateralized debt obligations (CDO), and many others, became practically unsalable. Another, current example is that it can be impossible to convert certain foreign currencies to any one of the major currencies.

It is the case, however, that the majority of assets on a given market at a given time are liquid enough to allow for small-volume trading. Therefore liquidity risk is often insignificant for small-sized traders. However, the impact of trading becomes significant as the volume grows, as may be the case for large institutional investors, and must be taken into account as an additional cost of trading.

The notion of \textit{price impact} is fundamental when it comes to large-volume trading.  Without such price pressure, all trading strategies would be infinitely scalable, since the cost would remain unchanged regardless of the size of the trader. In addition to mechanical liquidity consumption, price impact can be seen as an information game. If the market manages to guess a trader's intend to buy a large quantity of some asset (or correlated assets), they can be outrun by those traders who know and seek to benefit from that knowledge.

Understanding the determinants of impact is therefore crucial from several perspectives:
\begin{itemize}
    \item For the economist, modelling impact provides an understanding of how prices change and how they reflect the asset's fundamental value. This in turn requires to develop a micro-model for the statistics of prices;
    \item for the trader, price impact may represent a large fraction of execution costs. Assessing the impact of any trading strategy is of utmost importance to asset managers, since too much trading (whether in volume or in frequency) can deteriorate the performance of a strategy or turn a profitable strategy into a money-losing one;
    \item for the regulator, acknowledging the existence of impact means that fair-value accounting using mark-to-market prices is over-optimistic. A second important consequence is that excessive trading costs may impede execution and reduce market fluidity. Finally, price impact is an important bridge between market design and systemic risk prevention. Therefore, a better understanding of impact would also be helpful from the perspective of market microstructure regulation. 
\end{itemize}

\subsection{The Kyle approach}
The first price impact model is perhaps due to \cite{kyle1985continuous}. In modern terms, it postulates that impact is \textit{permanent} and \textit{linear} both in time and in the traded volume. 
A single trade of volume $q$ and sign $\epsilon \in \{ \pm 1 \}$ leads to a price move proportional to $\epsilon \cdot q$. This leads to total price change between times $0$ and $t$ equal to $\alpha \displaystyle{\sum_{s \leq t}} \epsilon_s q_s$, for some constant $\alpha.$
If the market price follows a random walk, then the signs of the trades $(\epsilon_s)$ should be uncorrelated.
However, real data shows that order signs are correlated and that this autocorrelation decays very slowly with time.
Furthermore, widespread empirical evidence 
\citep{bucci2019crossover,toth2016square, toth2011anomalous, torre1998market,loeb1983trading} 
indicates that the impact of trading nearly follows a square-root law in the volume traded, which can unfortunately not be accommodated within this model.

\subsection{The propagator approach}

Kyle's original proposal is closely related to \textit{propagator models}, which posit a time-decaying kernel for the impact of trades \citep{bouchaud2004fluctuations}. In these models total impact is the sum of the impact of individual trades: given a positive, non-increasing kernel $\mathcal{G}: \mathbb{R}_+ \to \mathbb{R}_+$, the impact at time $t$ of a series of trades $\left((q_s,\epsilon_s)\right)$ is given by $y_t = \sum_{s \leq t} \epsilon_s q_s \mathcal{G}(t-s).$ This can be stated equivalently in the continuous case by writing $y_t = \displaystyle{\int_0^t} \textnormal{ds}m_s \mathcal{G}(t-s)$, where the trading rate satisfies $\textnormal{d}(\epsilon q) = m \textnormal{ds}$. One may additionally posit that $\displaystyle{\lim_{t \to \infty}} \mathcal{G} = 0$, so that impact vanishes at longer timescales; see \cite{gueant2013permanent} for a discussion on transient and permanent impact. 

\cite{taranto2018linear} provide a comprehensive summary of propagator models. This family has the advantage of being simple to design and leading to tractable analytical results. Unfortunately, it does not provide any further explanation about the origin of impact.

\subsection{The Donier \textit{et al.} approach}
The complexity of today's financial markets stems from the many traders who continuously interact with one another to form prices; the limit order book forms the basis behind this interaction. Recently several order book models have been proposed that adopt an agent-based approach \citep{huang2019glosten, mastromatteo2014agent}. Such an approach goes beyond \textit{ad hoc} formulations and offers a glimpse into a phenomenon mostly viewed as a stylized statistical fact, which provides white-box understanding of the price formation process.

Salient in this category is \cite{donier2015fully}'s \textit{latent order book} model, which argues that the visible order book is insufficient to reflect true supply and demand. The fundamental reason behind this is the asymmetry between liquidity providers and liquidity takers, which has become widely accepted ever since the classical work of \cite{glosten1985bid}, and is closely related to the notion of adverse selection. In reality, the visible order book mostly displays the activity of high-frequency participants, whereas the intentions of low-frequency actors remain concealed up until the time immediately preceding execution.

\begin{figure}
    \centering
    \includegraphics[width=0.3\textwidth]{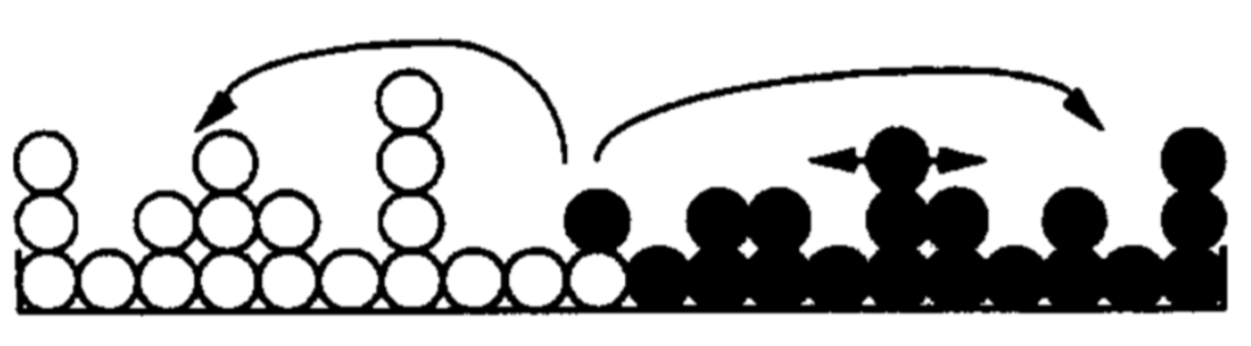}
    \caption{
    A white particle diffusing from the bid side collides and annihilates when it bumps into a black particle diffusing from the ask side \citep{bak1996price}. The annihilation corresponds to a real transaction taking place in the order book.
    }
    \label{bak}
\end{figure}

The authors posit \textit{linear} density in the neighborhood of the efficient price. In this model, impact is a consequence of two opposing phenomena:
\begin{itemize}
    \item liquidity consumption, which increases spread and the cost of trading;
    \item diffusion, which pushes the prices back towards a mean-reverting equilibrium, and is illustrated in Figure \ref{bak}.
\end{itemize}

The findings of the model have received empirical validation \citep{donier2015million} on one of the largest available trading datasets.
The authors analyze trading activity on a major Bitcoin exchange, demonstrating that the salient concavity characteristic of impact remains valid independently of market venue. 
Even when individual orders cannot be systematically detected (due to the anonymity enforced on trading venues), even in the absence of a notion of fundamental value (which makes little sense as of today on the Bitcoin market), the persistence of impact suggests that a robust self-organizing mechanism is at work.
This seems best explained by an agent-based approach which universally models trader behavior.

\subsection{Our contribution}
The primary motivation for our work is to enrich the minimal proposal of \cite{donier2015fully} with a conceptually simple yet practically important ingredient. 
Mean reversion is the assumption that the price tends to revert back towards its historical average, and is a well-documented phenomenon across many markets and venues \cite{palwasha2018speed,poterba1988mean,narayan2007mean}.
From the practitioner's perspective, this added ingredient provides an additional degree of freedom in the model, thereby increasing its expressivity and allowing better calibration on real data.

This paper explicitly analyzes how the mean-reversion of prices in the latent order book model affects impact, and shows that the latter is negatively related to the speed of price reversion.
Thereby this speed may be interpreted as another dimension of liquidity, in line with previous price impact models \citep{huberman2005optimal}. 
A salient feature of our model is that mean-reversion only acts in the medium to long time range, in accordance with empirical evidence indicating that it is not observed in the short term \citep{chakraborty2011market}. 

We also provide an existence theorem for the average order book density under impact.
This mathematical result also gives for free an existence result for \cite{donier2015fully}'s original model. Finally, we complement our theoretical analysis with a numerical scheme that simulates the entire order book and might be of independent interest.

\section{Model formulation}
The original latent order book model of \cite{donier2015fully} incorporates two key ingredients:

\begin{itemize}
    \item drift-diffusion, which models the random mechanical fluctuations in price and occurs at the rate $D \equiv \frac{\sigma^2}{2}$;
    \item order matching, which reflects the fact that a mutually beneficial agreement has been found whenever supply and demand are simultaneously nonzero. One generally works in the limit of infinitely reactive markets in order to simulate clearing of buy and sell orders.
\end{itemize}

Underlying this approach is an implicit mean-field assumption \citep{lasry2007mean} so that one interprets the order book densities as \textit{average} densities. 

Building on this purely diffusive motion, we allow the agents to adjust their current price further towards the underlying latent price of the asset. Our proposal maintains the indispensable diffusive behavior and enhances it with a mean-reverting component. 
Two additional ingredients, \textit{deposition} and \textit{cancellation}, are alluded to but not analyzed fully by \cite{donier2015fully}. While not key to our model, we explicitly study them for completeness in appendix A.

To obtain the new price dynamics we start by writing the microscopic evolution in a non-rigorous way before transforming it into a partial differential equation.

We write that each agent reassesses its price as follows:
\begin{equation}
    p_{i,t} \mapsto p_{i,t+dt} = p_{i,t} +\eta_{i,t} - \kappa \cdot (p_{i,t} - B_t) \textnormal{dt},
\end{equation}
where the process $(B_t)_t$ represents the \textit{reference price}, which could either be exogenous (such as a Brownian motion) or endogenous (such as taking the current market price and plugging it in, leading to a feedback loop). 
The noise variables $\eta_{i,t} \sim N(0,\sigma \textnormal{dt})$ are agent-dependent, so that $p_{i,t} + \eta_{i,t}$ represents agent $i$'s best estimate of the fundamental price. Finally; $\kappa > 0$ quantifies the return force towards $B_t$; its intuitive role is identical to that of the string constant in Hooke's law.

Starting from 
\begin{align*}
\begin{split}
    p_{i,t} &= \frac{1}{1-\kappa \textnormal{dt}} (p_{i,t+\textnormal{dt}} - \eta_{i,t}
    + \kappa B_t \textnormal{dt}) \\
    &\approx p_{i,t+\textnormal{dt}} - \eta_{i,t} + \kappa (p_{i,t+\textnormal{dt}} - B_t)\textnormal{dt},
\end{split}
\end{align*}

one performs a second-order expansion:

\begin{equation}
\label{meanreversion-PDE}
    \begin{split}
    \varphi(x,t+\textnormal{dt})
    &= \int \mathbb{P}(\eta) 
    \int \textnormal{dy} \delta(x-\eta+\kappa(x-B_t)\textnormal{dt}-y)  \\
    &\approx \varphi(x) + (0 + \kappa(x-B_t))\partial_x\varphi(x,t)\textnormal{dt}
    + \frac{\sigma^2}{2} \partial_{xx}\varphi(x,t)
    \end{split}
\end{equation}

so that the density of orders in the book evolves according to the partial differential equation:
\begin{equation}\label{main-PDE}
    \partial_t \varphi(x,t) = \kappa (x-B_t) \partial_x \varphi(x,t)
    + \frac{\sigma^2}{2} \partial_{xx} \varphi(x,t), \forall y \in \mathbb{R}, \forall t \geq 0.
\end{equation}

This means that agents reassess their price all the more as they are far from the reference price $B_t$, the intensity of reassessment being determined by the parameter $\kappa$.

These dynamics would not be complete without a boundary condition. We specifically consider the initial-time condition

\begin{equation}
    \varphi_{| t = 0} = \mathcal{L}y \textnormal{ on } \mathbb{R}.
\end{equation}

This condition, together with diffusion, ensures that the order book far from the current market price replenishes at the constant rate $\mathcal{L}$, so that $\partial_y \varphi (y,t) \xrightarrow[y \to \pm \infty]{} \mathcal{L} $. 
We refer the reader to \cite{donier2015fully} for a in-depth presentation of the original model.

\section{The shape of the order book}\label{analytical-resolution}
The dynamics \eqref{main-PDE} of the order book in presence of the reference price $B_t$ are no longer linear.
However, we show below that they become linear after a certain change of reference frame that centers the price.

To do so one performs the change of variable $y = x - f(t)$ where 
$f(t) = \kappa \int_0 ^t \textnormal{ds}B_s e^{-\kappa(t-s)}$. The function $f(t)$ may be seen as a weighted average of the reference price, since it is nearly equal to $
\frac{1}{\int_0^t \textnormal{ds} e^{\kappa s}} \int_0^t e^{\kappa s}dB_s
$. 
Here $\tau \equiv {\kappa}^{-1}$ defining the half-life averaging memory. 
The function $f$ satisfies the differential equation $f' + \kappa f = \kappa B_t$.
As a result, the translated function $\phi(y,t) \equiv \varphi(y + f(t),t)$ satisfies the \textit{linear} partial differential equation:

\begin{equation}\label{linear-pde}
    \partial_t \phi(y,t) = \kappa y \partial_y \phi(x,t) + \frac{\sigma^2}{2}\partial_{yy} \phi(y,t).
\end{equation}

\subsection{Approach by separation of variables}\label{approach-by-separation-of-variables}

Here one is interested in simple solutions to \eqref{linear-pde} of the form $\phi(y,t) = g(y)h(t)$. This leads to the system

\[
\systeme*{\kappa \cdot y \cdot g\prime (y) + \frac{\sigma^2}{2} g''(y) = c\cdot g(y), 
h'(t) = c \cdot h(t) }
\]
where $c$ is an \textit{a priori} arbitrary real constant. However, it turns out that $c=0$ is the only acceptable value, since at large times one does not expect the order book to collapse to $0$ or diverge to $+\infty$. This ensures \textit{de facto} that $h$ is constant, leading to stationary solutions of the form $\phi_{st}(y) = c g(y)$ where $g$ satisfies $\kappa y g\prime (y) + \frac{\sigma^2}{2} g''(y) = 0$. 
A straightforward calculation will yield $\phi_{st}(y) = c_0 + c_1 \int_{-\infty}^y \textnormal{dx} e^{-\frac{\kappa x^2}{\sigma^2}}$. 
\newline Going back to the original reference frame yields
\begin{equation*}
    \varphi_{st}(x,t) = c_0 + c_1 \int_{-\infty}^{x-f(t)}
    \frac{\textnormal{dy}}{\sqrt{2 \pi \sigma^2}}
    e^{-\frac{\kappa(y-f(t)^2}{2\sigma^2}}.
\end{equation*}
This family of stationary solutions is controlled by two parameters:
\begin{itemize}
    \item $c_0 = \displaystyle{\lim_{x \to -\infty}}
    \varphi_{st}(x,t)$, which determines the latent \textit{buy} volume far from the market price
    \item $c_1$, which reflects the latent \textit{sell} volume far from the market price since
    $\displaystyle{\lim_{x \to +\infty}} \varphi_{st}(x,t) = c_0 + c_1$.
\end{itemize}

\begin{figure}
    \centering
    \includegraphics[width=\textwidth]{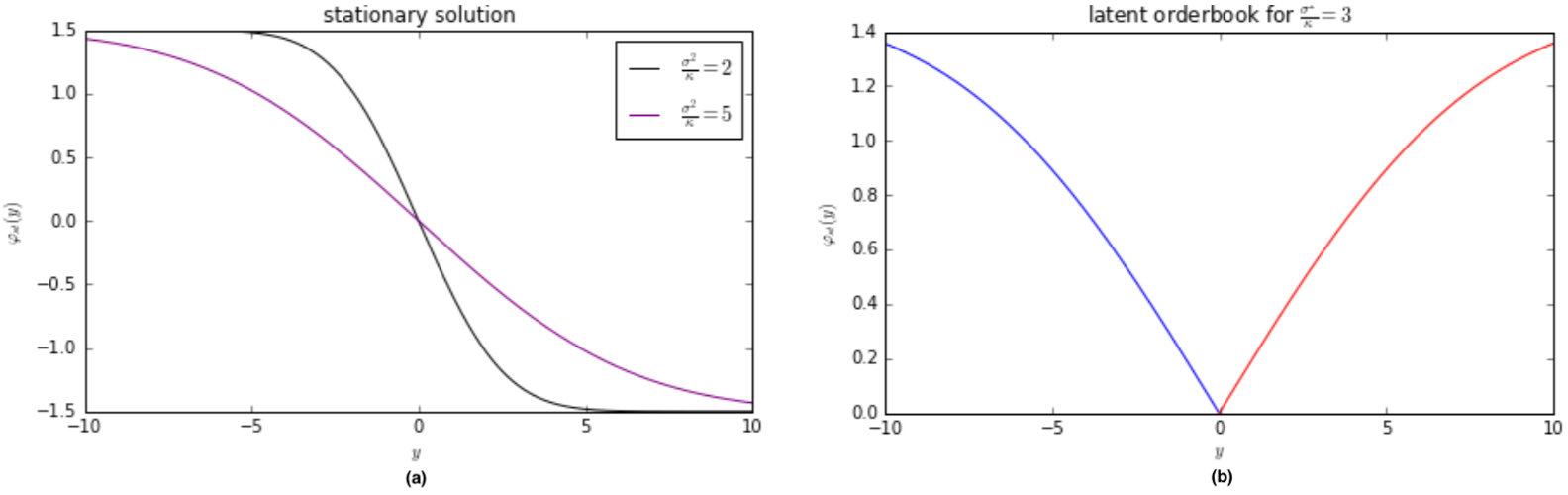}
    \caption{\textit{(a):} example of a stationary solution with $c_0 = 1.5, c_1 = -3 \sqrt{\frac{\kappa}{\pi \sigma^2}}, \frac{\sigma^2}{\kappa}\in \{2, 5 \}$.
    \newline
    \textit{(b):} the corresponding order book, where the blue curve represents the bid side and the red curve the ask side. In the new reference frame the equilibrium price is $0$.}
\end{figure}

\subsection{Full resolution}
The differential equation \eqref{main-PDE} is nontrivial to tackle directly due to the nonlinearity involving the reference price.

The initial change of variable $y = x-f(t)$ translated spatial coordinates to follow this reference price. 

In this section we consider the change of variable $y = e^{\kappa t}(x-f(t))$. The additional change in time scales allows to transform the price dynamics into a simple diffusion process (although at a time-dependent diffusion rate).

That is, by defining $\psi(y,t) = \varphi (e^{-\kappa t}y+f(t), t)$ the dynamics take the expression

\begin{align}\label{simple-pde}
    \partial_t \psi(y,t) = \frac{\sigma^2}{2}e^{2\kappa t} \partial_{yy}\psi(y,t).
\end{align}

We now proceed to show how to solve \eqref{simple-pde}, which is an application of Fourier calculus\footnote{To this end one may first recall the following elementary facts: 
\begin{itemize}
    \item given an integrable function $f \in L^1(\mathbb{R},\mathbb{C})$, its Fourier transform is defined by $\mathcal{F}(f): k \mapsto \int_{-\infty}^k \textnormal{dy}f(y) e^{-iky}$.
    \item If $\hat{f}\equiv F(f)\in L^1(\mathbb{R},\mathbb{C})$, in particular if $f$ is continuous, its inverse Fourier transform $\mathcal{F}^{-1}(\hat{f}):y \mapsto \frac{1}{2\pi} \int_{\infty}^{+\infty}\textnormal{dk}f(k)e^{+iky}$ is well-defined and satisfies $\mathbb{F}^{-1}(\hat{f}) = f$ almost surely.
\end{itemize}
}.

Applying the Fourier transform to the space variable $y$ and using the fact that $\mathcal{F}[\partial_{yy}\psi] = -k^2 \psi.$ leads to the PDE

\begin{align*}
    \partial_t \hat \psi(k,t) = - \frac{\sigma ^2}{2}e^{2\kappa t}k^2\hat \psi(k,t).
\end{align*}

The solution to this last equation can be determined by variation of constants. One readily obtains
$\hat \psi(k,t) = g(k) e^{-\frac{C_\kappa(0,t)\sigma^2k^2}{2}},$ where $g = \mathcal{F}[\psi_0]$ and the initial condition
$\psi_0 = \varphi_0 \equiv \varphi(\cdot,t=0)$ is continuous over $\mathbb{R}$. We also define

\begin{align}\label{definition-C}
    C_\kappa(s,t) \equiv \int_s^t \textnormal{du}e^{2\kappa u}
    = \frac{1}{2\kappa}(e^{2\kappa t} - e^{2\kappa s})
\end{align}

for $0 \leq s \leq t.$

The final solution is obtained by going back to the space domain. The transform of the product of two
functions is the convolution of their transforms, hence

\begin{align*}
    \psi(y,t) = \int_{-\infty}^{+\infty} \frac{\textnormal{du}}{\sqrt{2\pi C_\kappa(0,t)\sigma^2}} \varphi_0(u)e^{-\frac{(u-y)^2}{2C_\kappa(0,t)\sigma^2}}.
\end{align*}

This solution is well-defined for a wide spectrum of initial conditions (essentially functions with sub-exponential growth) so long as the integrand remains in $L^1$ for all times $t\geq 0$. This includes in particular linear functions.

The diffusion rate, defined by the variance of the heat kernel, increases exponentially with time. 
We conclude by noting that one indeed recovers the usual impact profile when $\kappa \to 0.$

\section{Price impact}
Assume that a large trader, such as an institutional investor, executes a large order according to the schedule $(m_t)_{t \in [0,T]}$.
We will show that that the impacted price satisfies the integral equation
\begin{align} \label{impacted-price}
    y_t = \frac{1}{\mathcal{L}}\int_0 ^t 
    \frac {\textnormal{ds}m_s e^{\kappa s}} {\sqrt{2\pi\sigma^2 C_\kappa(s,t)}} e^{-\frac{(y_t-y_s)^2}{2\sigma^2C_\kappa(s,t)}}.
\end{align}

One can first observe that this impact profile generalizes that of \cite{donier2015fully}, which took the slightly simpler expression 
\begin{align}\label{donier-impact}
    y_t = \frac{1}{\mathcal{L}} \int_0^t
\frac{\textnormal{ds}m_s} {\sqrt{2\pi \sigma^2 (t-s)}}
e^{-\frac{(y_t-y_s)^2}{2\sigma^2(t-s)}}.
\end{align}

Further comparisons are provided in sections \ref{small-trading-rates} and \ref{existence}.

\subsection{Impact profile}
Starting from the original dynamics \eqref{main-PDE}, one introduces a metaorder $(m_t) \in \mathcal{C}([0,T])$ executed at the market price $x_t$:

\begin{align}\label{impacted-price-equation}
    \partial_t \varphi(x,t) = \kappa (x-B_t) \partial_x \varphi(x,t)
    + \frac{1}{2} \sigma^2 \partial_{xx} \varphi(x,t) + m_t \delta(x-x_t),
\end{align}

where $\delta$ denotes the Dirac delta function.

This means that in the new reference frame we have:

\begin{align*}
    \partial_t \psi(y,t) = \frac{\sigma^2}{2}e^{2\kappa t} \partial_{yy}\psi(y,t)
    + m_t \delta(e^{-\kappa t}y - x_t).
\end{align*}

Since the Dirac distribution $\xi \mapsto \delta(\xi)$ is homogeneous of degree $-1$ with respect to $\xi$, we may rewrite this equation as

\begin{align*}
    \partial_t \psi(y,t) = \frac{\sigma^2}{2}e^{2\kappa t} \partial_{yy}\psi(y,t)
    + m_t e^{\kappa t} \delta(y - y_t),
\end{align*}
where $y_t \equiv e^{\kappa t}(x_t - f(t))$ is the reframed market price, \textit{i.e.} the zero of $\psi(\cdot,t)$ (whose existence and unicity depend on the initial condition, although they will become clear at the end of this paragraph). One also observes that the metaorder volume becomes $m_t e^{\kappa t}$ instead of $m_t$. This is a natural consequence of our change of variable which made the new space variable depend on time.

Moving to the Fourier domain in space gives 
\begin{align*}
    \partial_t \hat \psi (k,t) = - \frac{\sigma ^2}{2}e^{2\kappa t}k^2 \hat \psi(k,t)
    + m_t e^{\kappa t - iky_t}.
\end{align*}

This is solved by variation of constants and gives

\begin{align*}
    \hat\psi(k,t) = g(k)e^{-\frac{\sigma^2C_\kappa(0,t)k^2}{2}}
    + \int_0^t \textnormal{ds}m_s e^{\kappa s -iky_s -\frac{\sigma^2 k^2 C_\kappa(s,t)}{2}},
\end{align*}

where $g$ denotes the Fourier transform of the initial condition.

Applying the inverse Fourier transform therefore leads to

\begin{align*}
    \psi(y,t) = \frac{1}{\sqrt{2\pi C_\kappa(0,t)\sigma^2}}(\psi_0 * e^{-\frac{x^2}{2C_\kappa(0,t)\sigma^2}})(y)
    + \int_0^t\textnormal{ds}m_se^{\kappa s}
    \mathcal{F}^{-1}[e^{-iky_s-\frac{\sigma^2C_\kappa(s,t)k^2}{2}}](y).
\end{align*}

For a linear initial condition $\psi_0(y)=\varphi_0(y)=-\mathcal{L}y$, this gives

\begin{align*}
    \psi(y,t) = -\mathcal{L}y + \int_0^t\frac{\textnormal{ds}m_se^{\kappa s}}{\sqrt{2\pi C_\kappa(s,t)\sigma^2}} e^{-\frac{(y-y_s)^2}{2\sigma^2C_\kappa(s,t)}}.
\end{align*},
which immediately recovers the desired impact profile of \eqref{impacted-price}.

\subsection{Small trading rates}\label{small-trading-rates}
In today's fragmented markets, the trading activity of any individual agent, although large, is generally an order of magnitude smaller than the total market activity. It is therefore especially interesting to study the simple special case of small trading rates.

This regime is defined by $\lVert m \rVert \ll L\sigma$ and allows to make the approximation of small impacts $(y_t - y_s)^2 \ll C_\kappa(s,t)$. This leads to

\begin{align} \label{small-trading-rates-eq}
    y_t = \frac{1}{\mathcal{L}} \int_0 ^t \frac{\textnormal{ds}e^{\kappa s}}{\sqrt{2\pi C_\kappa(s,t)\sigma^2}}m_s.
\end{align}

Therefore impact becomes \textit{linear} and falls broadly within the family of propagator models that was introduced in section \ref{introduction}.
One may recall that in \cite{donier2015fully}'s original proposal, impact for small trading rates was given by
\begin{align*}
y_t^{|(\kappa = 0)} = 
\frac{1}{\mathcal{L}} \int_0 ^t \frac{\textnormal{ds}}{\sqrt{2\pi (t-s) \sigma^2}}m_s,
\end{align*}

corresponding to the limit $\kappa \to 0$ in \eqref{small-trading-rates-eq}). Now mean-reversion counters this square-root decay kernel, nuancing the growth of impact. More precisely, one can establish that $y_t \leq y_t^{|(\kappa = 0)}$ and this is a consequence of the inequality $\frac{e^{\kappa s}}{C(s,t)} \leq \frac{1}{t-s}$.

For an additional bit of insight, let us consider another simple regime, that of constant trading rates $m_t = m_0, \forall t \in [0,T]$.

A straightforward calculation using the change of variable $v = e^{-\kappa (t-s)}$ confirms \textbf{concave impact}:

\begin{align*}
    y_t = \frac{m_0}{\mathcal{L}\sigma \sqrt {\kappa \pi}}
    \int_{e^{-\kappa t}}^1\frac{\textnormal{dv}}{\sqrt{1-v^2}}
    = \frac{1}{\mathcal{L}\sigma \sqrt{\kappa\pi}}\left( \frac{\pi}{2} - \arcsin(e^{-\kappa t})\right).
\end{align*}

Since $\arcsin x \isEquivTo{x\to 1} \frac{\pi}{2} - \sqrt{2(1-x)}$, impact at shorter time scales is roughly
\begin{align*}
 y_t \isEquivTo{t \to 0} \frac{m_0}{\mathcal{L}\sigma \sqrt {\kappa \pi}}\sqrt{2t},
\end{align*}
and one \textbf{recovers the original square-root law}.

At larger time scales, however, impact converges to a finite nonzero value:
\begin{align*}
    y_t \displaystyle{\to_{t \to +\infty}} \frac{\pi m_0}{\mathcal{L}\sigma \sqrt{2\kappa}},
\end{align*}
 in sharp contrast with the divergence of impact in \cite{donier2015fully}'s proposal. This is explained in simple terms by observing that agents continually reassess their price towards the fundamental value, thereby providing resistance against price increases by means of added liquidity.

\begin{figure}
    \centering
    \includegraphics[width=0.4\textwidth]{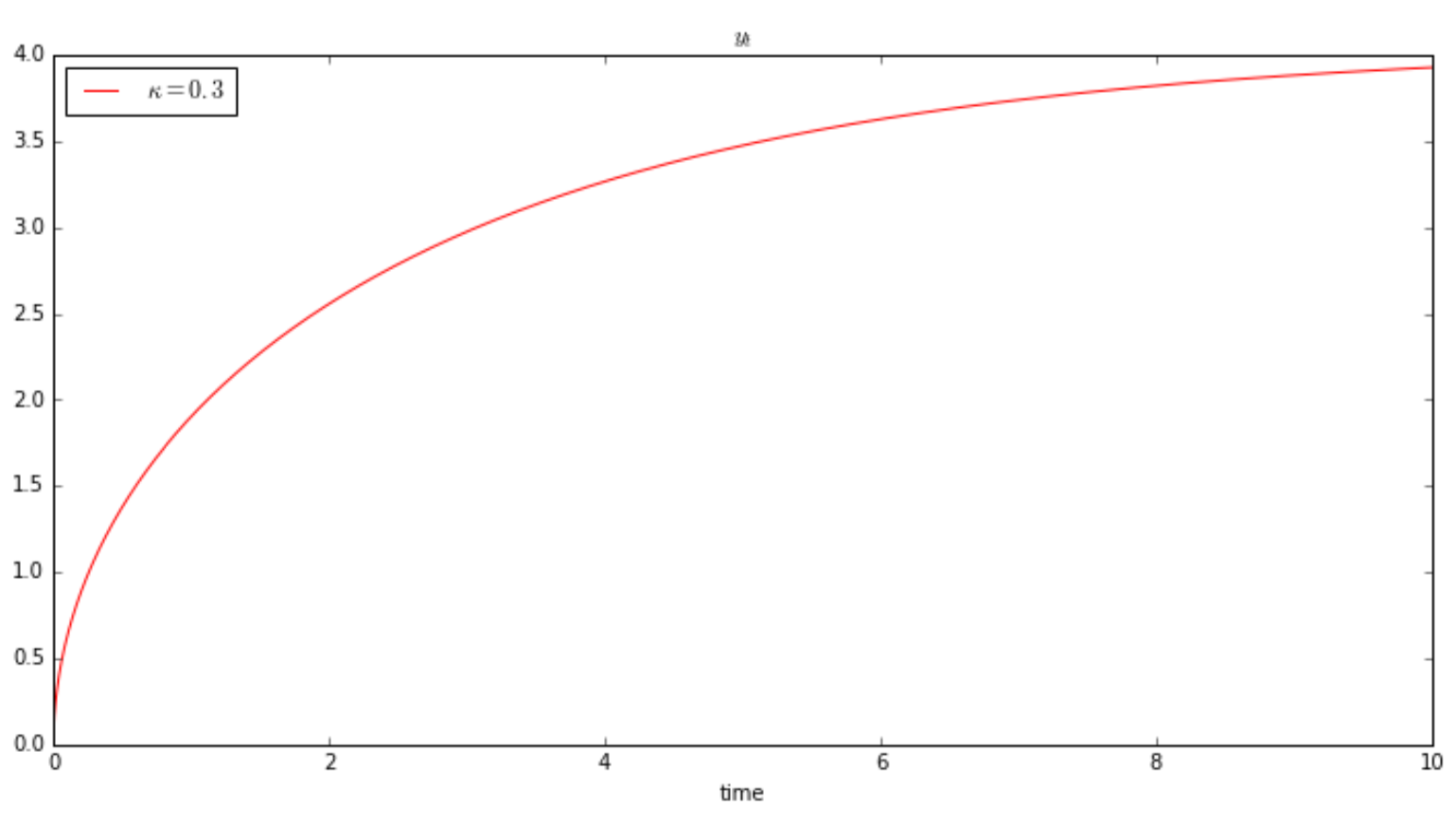}
    \caption{Impact profile in the limit of small trading rates. We recover a linear propagator that leads to concave impact. The propagator replicates the square-root behavior at shorter times, and converges to a finite value at larger times.}
\end{figure}

\subsection{Existence and uniqueness results}\label{existence}
There is, to the best of our knowledge, no closed-form solution to the impacted price equation \eqref{impacted-price}. Fortunately, we still have the

\begin{theorem}
\label{thm:existence}
Let $m:[0,T] \to \mathbb{R}$ be any continuous execution strategy. 
Then the impacted price equation \eqref{impacted-price} admits a solution $y$ defined over $[0,+\infty]$. Furthermore any solution is infinitely differentiable on its domain.
\end{theorem}

The proof is given in Appendix B.
As a corollary of Theorem \ref{thm:existence}, one gets for free a similar existence result for the original proposal of \cite{donier2015fully}.
This is because $C_\kappa (s,t) \xrightarrow[\kappa \to 0]{}t-s$and this is sufficient for the proof to remain valid.

\begin{corollary}
Let $m: [0,T] \to \mathbb{R} $ be any continuous execution strategy. Then the impacted price equation \eqref{donier-impact} admits an infinitely differentiable solution defined over $[0,+\infty]$.
\end{corollary}

\textit{Possible extensions:} does the impacted price admit a \textit{unique} solution? Intuitively we expect the answer to be positive in light of the mean-field assumption, which interprets the densities as average densities.

\section{Diffusion and mean-reversion}\label{diffusion-and-mean-reversion}
It was seen in section \ref{analytical-resolution} that the market price tracks a re-weighted average $f(t)$ of the reference price $B_t$, where the re-weighting occurs over a rolling time window of width $\tau = \kappa^{-1}$ and the tracking intensity is determined by $\kappa$. This section makes explicit the influence of the microscopic parameters governing the price dynamics \eqref{main-PDE}, namely the volatility $\sigma$ and the mean-reversion intensity $\kappa$. Their effect will be illustrated analytically both on the impacted price and \textit{market mispricing}, a measure of deviance from equilibrium. Further numerical comparisons are provided in section \ref{numerical-experiments}.

\subsection{Effect on the impacted price}
To get a first intuition on diffusion, we consider the stationary solution of section \ref{approach-by-separation-of-variables}. One sees that an increase in volatility leads to a increase in liquidity near the market price, making the market more robust to small perturbations, and thereby reducing impact. A similar conclusion can be derived analytically from \eqref{small-trading-rates-eq} in the case of small trading rates. This added robustness is of course expected since the diffusive jumps (illustrated in Figure \ref{bak}) act as a smoothing mechanism, and tend to occur more frequently under increased variance.

The mean-reversion parameter $\kappa$ has the inverse effect.
A higher value increases the drive towards the market price, as can be seen either \textit{(i)} analytically by differentiating equation \eqref{impacted-price} with respect to $\kappa$; or \textit{(ii)} directly in equation \eqref{small-trading-rates-eq} in the special case of small trading rates; or \textit{(iii)} visually in Figure \ref{metaorder-execution}.

Finally, the limit $\kappa \to 0$ allows to recover the original impact profile of \cite{donier2015fully}, for which we have seen in section \ref{small-trading-rates} that $y_t \leq y_t^{|(\kappa = 0)}$.

\subsection{Effect on mispricing}\label{effect-on-mispricing}

Mispricing is defined as the difference between $B_t$ and $f(t)$ and quantifies how close the market price is to the underlying efficient price. Large mispricing values indicate market instability and a potential departure from equilibrium.

It can be seen through integration by parts (where the Ito component cancels since the integrand is a deterministic function of time) that

\begin{align*}
    f(t) = \kappa \int_0^t \textnormal{ds}e^{-\kappa (t-s)}B_s
    = \int_0^t \textnormal{d}(e^{-\kappa (t-s)})B_s
    = B_t - \int_0^te^{-\kappa(t-s)}dB_s,
\end{align*}
and thus
\begin{align}\label{mispricing-eq}
    B_t - f(t) = \int e^{-\kappa (t-s)} dB_s.
\end{align}
Therefore mispricing is a centered Gaussian variable with variance $\nu(\kappa,t) = \int_0^t e^{-2\kappa(t-s)}ds$, which decreases with $\kappa$. Hence the "convergence" of the impacted price to the reference price. This convergence underlines a notion of market stability in the sense that \textbf{mean-reversion reduces mispricing}.

\begin{figure}
    \centering
    \includegraphics[width=0.4\textwidth]{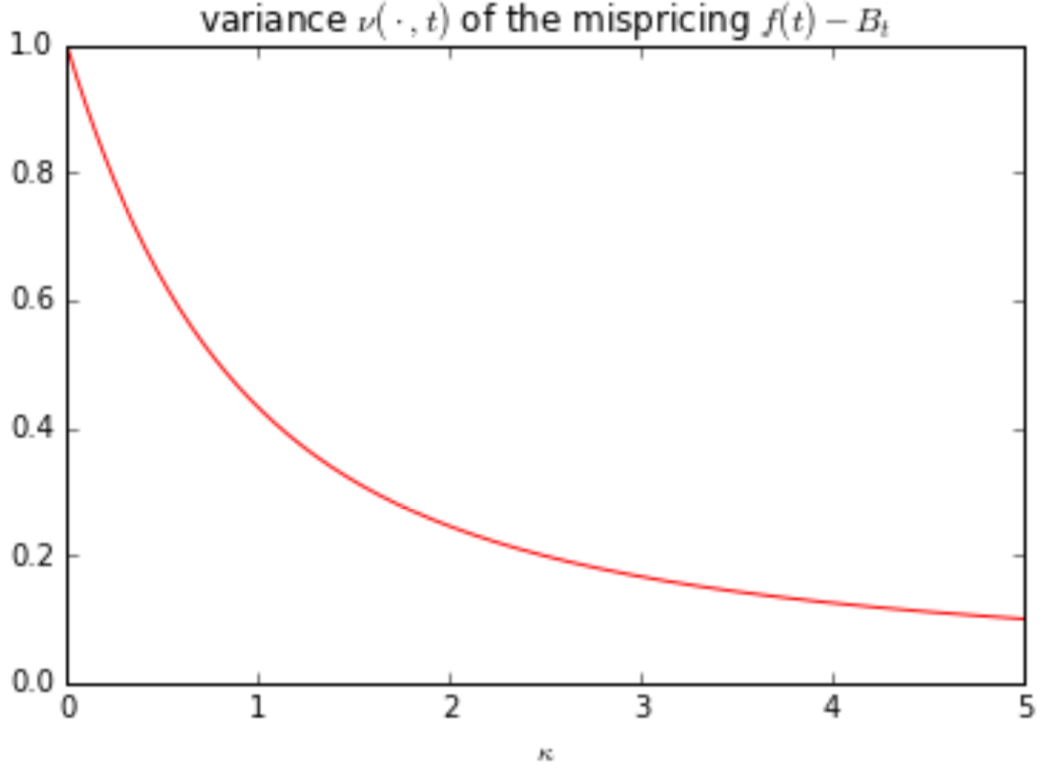}
    \caption{A strongly mean-reverting market is generally less mispriced. This is illustrated by the decay of variance of the mispricing variable defined in equation \eqref{mispricing-eq} with respect to the mean-reversion parameter $\kappa$ (displayed here at the time $t=1$).}
    \label{fig:stationary}
\end{figure}

\section{Numerical experiments}\label{numerical-experiments}

\begin{figure}
    \centering
    \includegraphics[width=0.85\textwidth]{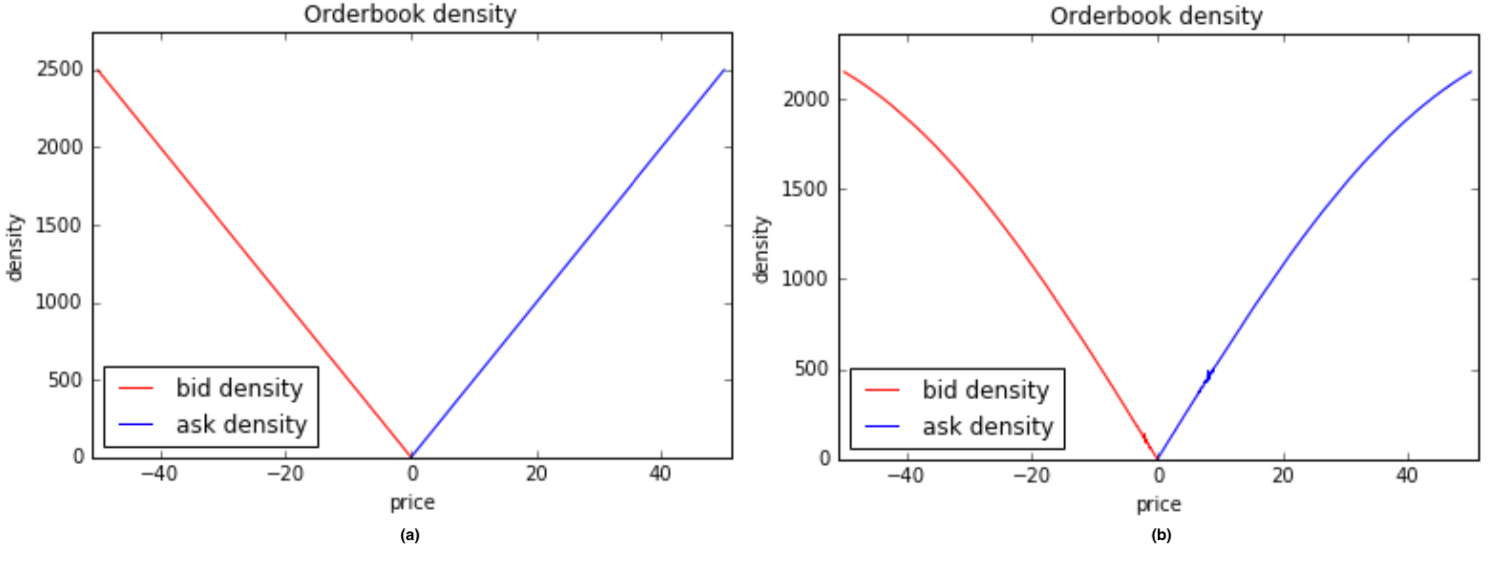}
    \caption{\textit{(a):} Initial (linear) shape of the order book $\varphi(y,0) = -Ly$, with $L=50$. \newline
\textit{(b):} Starting from this initial condition, $1500$ iterations of the Crank-Nicolson scheme are performed and the final order book shape is shown. The resulting density is in line with the stationary shape predicted in section \eqref{approach-by-separation-of-variables}.}
    \label{stationary-orderbook-density}
\end{figure}

This section sheds some numerical light on the predictions of the latent order book model. The dynamics \eqref{main-PDE} are simulated using a Crank-Nicolson finite-difference scheme.  Figure \ref{stationary-orderbook-density} illustrates the initial (linear) and resulting (nonlinear) shapes of the order book. Prices are restricted to finite support ranging from $-M$ to $M$ and covered by a regular grid with step size $\Delta x$. This means that the order book at time $t$ is therefore represented by a vector $X_t$ in $\mathbb{R}^{d}$, with $d = \frac{2M}{\Delta x} + 1$. In our simulations we take $M = 50$ and $\Delta x= 10^{-2}$, although in a real market $\Delta x$ could be the tick size and $M$ could be any sufficiently large value depending on the asset.

We use the first and second-order differentiation operators given by $A = -J + T$ and $B = -2 J + T + T^T$ respectively. Here $J$ denotes the diagonal-$1$ matrix and $T$ the upper-diagonal $1$ matrix, except that the first and last rows of $A$ and $B$ have been zeroed out to enforce reflective Dirichlet boundary conditions:
\begin{center}
$
\varphi (-M,t) = -\mathcal{L}\cdot M, \varphi (M,t) = \mathcal{L}\cdot M
$ and $
\partial_x \varphi (-M,t) = \partial_x \varphi (M,t) = 0, \forall t \geq 0,
$    
\end{center}
where $\mathcal{L}$ is the slope of the initial (linear) order book.

The discretized dynamics of \eqref{main-PDE} read:

\begin{equation}\label{discretization}
    \left( I - \frac{\sigma^2 \Delta T}{4 (\Delta x)^2}B \right) X_{t+1} =
    \left(
    I + \frac{\sigma^2 \Delta T}{4 \Delta x ^2}B
    + \kappa \frac{\Delta T}{\Delta X}(U - B_t \mathbbm{1})^T A)
    \right) X_t,
\end{equation}

where $I$ is the $d$-identity matrix, $\mathbbm{1}$ is the $d$-identity vector, and $
U = 
 \begin{bmatrix}
           -M \\
           -M + \Delta x \\
           \vdots \\
           M - \Delta x \\
           M
         \end{bmatrix}
         \in \mathbb{R}^d
$ is the grid vector.

Iterating this scheme allows to simulate the shape of the order book at any point in time. A finite time horizon $T = 1500$ seconds is considered together with a time step $\Delta T = 1$ second. The value of the market price is then deduced as the point of zero density.

\subsection{Mean-reversion and diffusion}
\begin{figure}
    \centering
    \includegraphics[width=\textwidth]{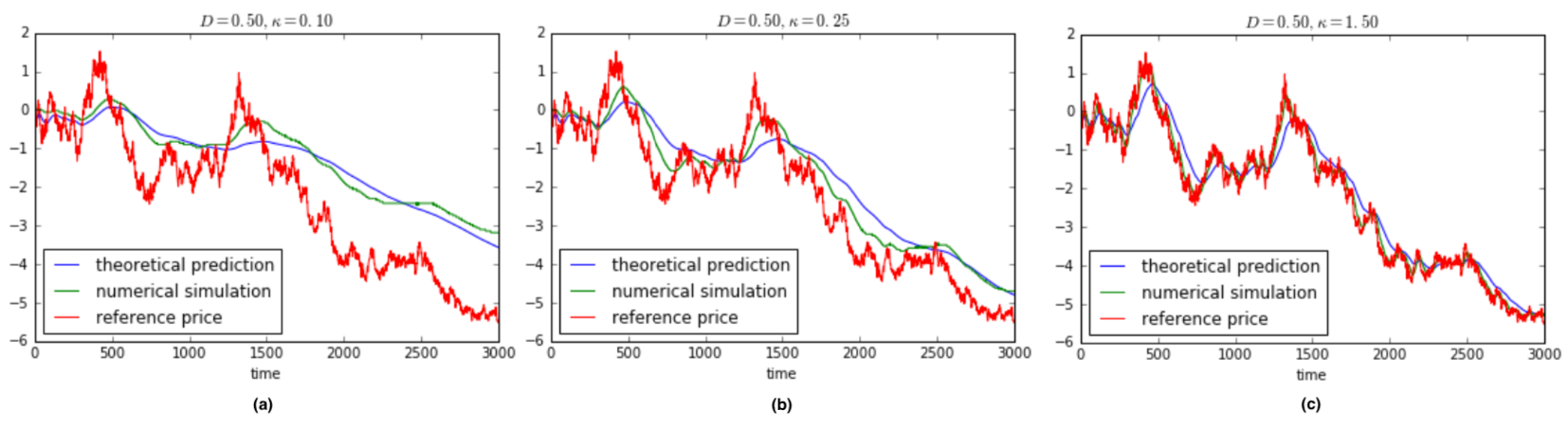}
    \caption{The reference price $(B_t)$, the market price simulated by finite difference method and the theoretical prediction $f(t)$ for $D \equiv \frac{\sigma^2}{2} = 0.5$ and \textit{(a):} $\kappa = 0.1$; \textit{(b):} $\kappa = 0.25$; and \textit{(c):} $\kappa = 1.5$. We observe a very close fit for larger values of $\kappa$, in line with the analysis of mispricing conducted in Section \eqref{effect-on-mispricing}.}
    \label{comparisons-kappa}
\end{figure}

\begin{figure}
    \centering
    \includegraphics[width=0.5\textwidth]{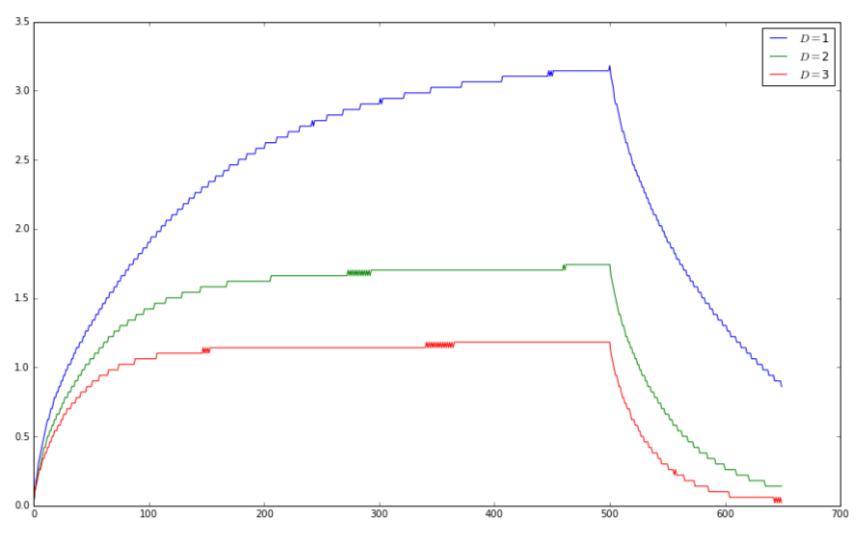}
    \caption{Evolution of the impacted price for the constant buy meta-order of Figure \ref{evolution-price}, executed during the first $T = 500$ seconds, which shows the mitigating effect of increased volatility on impact. 
    Here $\kappa = 1$ and $D \equiv \frac{\sigma^2}{2}$.}
    \label{comparisons-sigma}
\end{figure}

The influence of the mean-reversion and volatility parameters $\kappa$ and $\sigma$ was explored in Section \ref{diffusion-and-mean-reversion}; we illustrate them further in Figures \ref{comparisons-kappa} and \ref{comparisons-sigma}.

\subsection{The impacted price}
\begin{figure}
    \centering
    \includegraphics[width=\textwidth]{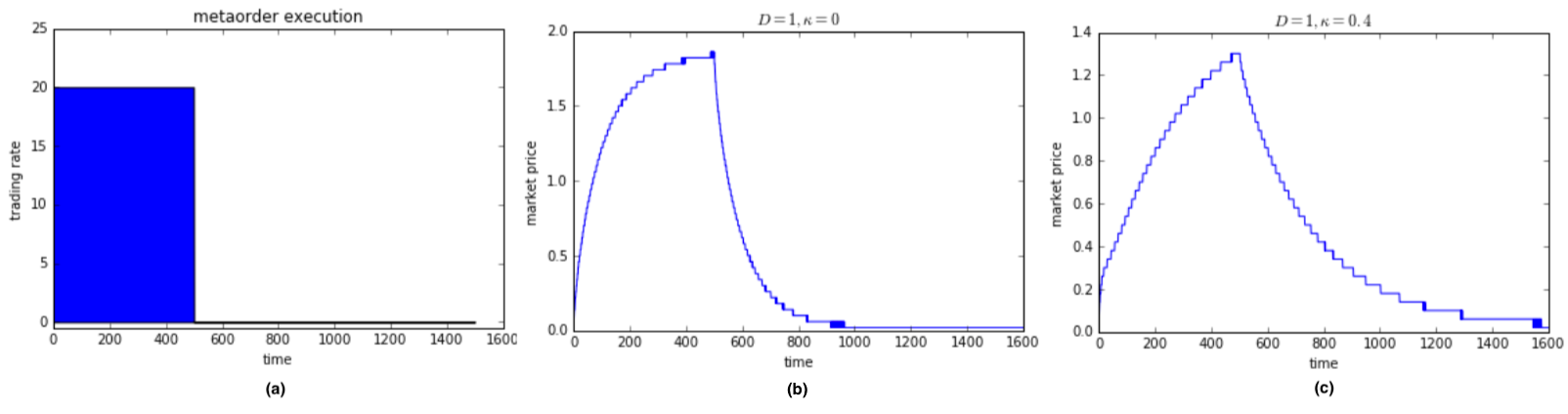}
    \caption{Evolution of the market price during and immediately following a buy metaorder executed over the first $T = 500$ seconds. 
    \newline \textit{(a):} meta-order execution profile
    \newline \textit{(b):} impacted price in the absence of mean-reversion
    \newline \textit{(c):} impacted price with mean reversion (after centering the price around the reference price). The mean-reverted curve provides a smoother profile where price impact is moderated by the pull towards the reference price.}
    \label{metaorder-execution}
\end{figure}

This simulation scheme conveniently allows to incorporate a metaorder.
A buy order is achieved by consuming the corresponding volume from the best available bid. If the buy order volume exceeds the best available volume, the remaining quantity is consumed from the next best bid, and so on until the entire order has been satisfied. In addition to Figure \eqref{metaorder-execution}, we have made available online a short video simulating the execution of that meta-order \footnote{https://www.youtube.com/watch?v=5QAqzERE5-g}.

\begin{figure}
    \centering
    \includegraphics[width=0.9\textwidth]{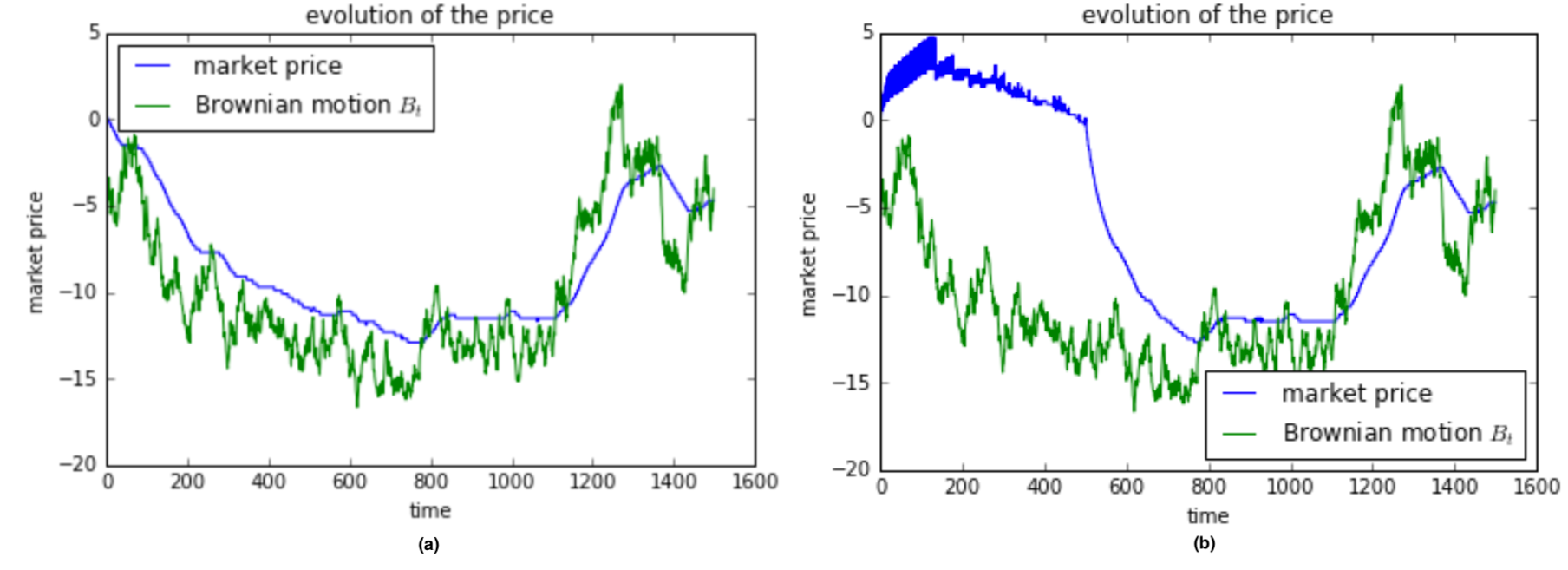}
    \caption{\textit{(a):} market price in the absence of meta-orders. For sufficiently strong mean reversion (simulated here with $\kappa = 0.5$), the market price closely tracks the reference price (simulated here with an affine Brownian motion).
    \newline \textit{(b):} market price in presence of the meta-order of Figure \ref{metaorder-execution}. One witnesses the trade-off between the positive push of the meta-order and the negative pull towards the reference price. The meta-order's effect dominates for the first $500$ seconds, but then its execution is complete and the market price reverts towards negative territory.}
    \label{evolution-price}
\end{figure}

We conclude this section with another interesting simulation, illustrated in Figure \ref{evolution-price}. It consists in taking an exogeneous reference price $B_t$ that opposes the direction of the meta-order. For a buy meta-order, this is achieved by adding a negative-drift component to the standard Brownian motion.  This provides the opportunity to observe clearly the trade-off between mean-reversion and price impact.

\clearpage
\section*{Discussion}
Our work extends the seminal proposal of \cite{donier2015fully}.
While the original dynamics were purely diffusive, this extension allows a mean-reverting behavior towards the "efficient" price, as is often observed in real markets. This modification represents one step towards a fully rational agent model, in contrast with \citeauthor{farmer2005predictive}'s zero-intelligence model.

On the other hand, our model reflects uninformed trading only, so that impact is a temporary statistical effect due to order flow fluctuations and liquidity imbalance.
Underlying this temporary nature is a key aspect of market structure, namely the difference between short-term and long-term term supply.
If a trader speeds up his buy trades, he depletes the short-term supply and increases the immediate cost for additional trades.
As more time elapses, supply gradually recovers and the price witnesses a mean-reversion to its initial value.

However, suppose that certain skillful agents can forecast short term price movements accurately.
For instance, if the agent correctly predicted (or was otherwise informed) that the price is about to rise, he is more likely to buy as an anticipation of this movement.
This should result in measurable correlation between trades and price changes, even if the trades by themselves have absolutely \textit{no effect} on the prices. 
Thereby the information processed by investors leads to permanent price moves reflecting a change in the asset's fundamental value.
This vision of price impact is purely based on information and is not captured by the present model, although it would be very desirable to do so.
One way towards this is to posit a joint distribution for the random drift together with trading volume, thereby allowing different meta-orders to interact in the order book. We leave this extension as a promising direction for future work.

For an adept of the mechanical vision, permanent impact is seen as the accumulation over time of the mechanical effects. 
For an adept of the informational vision, mechanical impact is a noise that reflects the activity of uninformed traders. 
This distinction gives a double interpretation of impact: on the one hand, market impact is a friction, and on the other it is the process by which prices adjust to new information.

\subsection*{Acknowledgements}
The author would like to thank Pierre Laffitte for giving him the opportunity to work on this subject and for fruitful mentoring. He also would like to thank Jiatu Cai for many interesting discussions. Thanks to the referee for numerous suggestions which helped to clarify the exposition and argumentation.

\newpage
\bibliographystyle{rusnat}
\bibliography{bib}

\newpage
\setcounter{secnumdepth}{0}
\section{Appendix A - Dynamics under deposition and cancellation}
The original proposal of \cite{donier2015fully} neglected deposition and cancellation of orders in order to derive the main differential equation \eqref{main-PDE}. We re-integrate these two parameters here for the sake of completeness and deduce the full dynamics of the latent order book as well as the impacted price. 

The differential equation now reads:
\begin{equation}
    \label{deposition-and-cancellation-eq}
    \partial_t \varphi(y,t) = -\nu \varphi(y,t) + \frac{\sigma^2}{2} \partial_y^2 \varphi(y,t)
    +\lambda \textnormal{sign}(y) + m_t \delta (y - y_t).    
\end{equation}

The cancellation rate $\nu$ leads to the notion of \textit{memory} of the order book, defined as $\tau =\nu^{-1}$. For times much larger than $\tau$, all orders in the book have been cancelled and replaced by new ones, so that its memory can be considered wiped. This time scale is of crucial importance because concave impact can only hold in the opposite regime where the metaorder duration satisfies $T \ll \tau$, so that the order book retains the information about the metaorder being executed. The reduction to the limit $\nu \to 0$ is therefore justified in that it ensures that execution time is negligible compared to the order book memory. Since deposition must not exceed cancellation too much in a balanced order book, one should also take the limit $\lambda \to 0$.

 The differential equation \eqref{deposition-and-cancellation-eq} can be solved explicitly using Fourier analysis similarly to Section \ref{analytical-resolution}. We let $g(y,s) =
e^{-\nu s} \left( \lambda \textnormal{sign}(y) + m_s \delta(y-y_s) \right)$
and $K(y,t) = \frac{e^{-\nu t -\frac{y^2}{2\sigma^2 t}}}{\sqrt{2\pi \sigma^2 t}} $, so that the solution obtained by convolution is

\begin{equation*}
    \varphi (y,t) = \left( \varphi_0 \star K(\cdot, t) \right) (y) +
    \int_0^t \left( g(\cdot,s) \star K(\cdot, t-s) \right) (y) \textnormal{ds},
\end{equation*}

where $\varphi_0$ is the initial order book at time $t=0$, namely $\varphi_0 (y) = -\mathcal{L}\cdot y$. This yields

\begin{equation*}
    \varphi(y,t) = -\mathcal{L}ye^{-\nu t} + \bigintss_0^t\textnormal{ds}e^{-\nu (t-s)} \left(
    \frac
    {m_s e^{- \frac{(y-y_s)^2}{\sqrt{2\pi \sigma^2(t-s)}}}}
    {\sqrt{2\pi \sigma^2 (t-s)}}
    + \lambda \phi (\frac{y}{\sigma \sqrt{t-s}})
    \right),
\end{equation*}
where $\phi (x) \equiv \mathbb{P}(|Z| \leq |x|)$ for Z a standard normal random variable.

Hence the impacted price satisfies the integral equation

\begin{equation*}
    y_t = \frac{1}{\mathcal{L}}
    \bigintss_0^t\textnormal{ds} \left(
    \frac{m_s e^{\nu s - \frac{(y_t - y_s)^2}{2\sigma^2(t-s)}}}
    {\sqrt{2\pi \sigma^2 (t-s)}}
    + \lambda \phi(\frac{y_t}{\sigma \sqrt{t-s}})
    \right).
\end{equation*}

This impact profile comes with the following interpretation:
\begin{itemize}
    \item The cancellation rate $\nu$ leads to a reparametrized trading rate $(m_s e^{\nu s})_s$ instead of $(m_s)_s$.
    In particular, the analytical approximations derived in section \ref{small-trading-rates}, including the square-root law, remain valid in several regimes so long as the small-trading assumption applies to the reparametrized rate as well.
    \item Impact is a decreasing function of the deposition rate $\lambda$. This can be verified analytically up to a 
    first approximation, or using the numerical scheme of section \ref{numerical-experiments}. This is of course expected as deposition contributes to the replenishment of available liquidity in the order book.
\end{itemize}

It is also straightforward to modify the numerical scheme \eqref{discretization} in order to include these two parameters, which can potentially increase the flexibility and expressive power of the model and provide a better fit when calibrated on real data.

\newpage
\section{Appendix B - Proof of Theorem 1}
\begin{proof}

 We work in the vector space $E \equiv (\mathcal{C}(I,\mathbb{R}, \lVert \cdot \rVert _\infty$ of real-valued continuous functions on $I \equiv [0,T]$. Fix $m = (m_t)_{0 \leq t \leq T}$ a function of $E$. For any $\epsilon > 0$, consider the function
$F_\epsilon: E \times E \to E$ defined by

\begin{align*}
    F_\epsilon(x,y)(t) = \int_0 ^{t-\epsilon}
    \frac{\textnormal{ds}m_se^{\kappa s}}{\sqrt{2\pi C_\kappa(s,t)}}
    e^{-\frac{(x_t-y_s)^2}{2C_\kappa(s,t)}}
\end{align*}

where $C_\kappa(s,t)$ is as in \eqref{definition-C}. The volatility is taken as $\sigma ^ 2 = 1$ here without loss of generality.

The first step is to show that the partial function $F_0(x,\cdot)$ has a fixed point $y(x)$ for any element $x\in E$. To do so we use Banach's fixed point theorem and show that for all $\epsilon >0$, $F_\epsilon(x,\cdot)$ has a unique fixed point $y_\epsilon(x)$. Then we prove that $y_\epsilon(x)$ converges to a fixed point of $F_0(x,\cdot)$ as $\epsilon \to 0$.

Given $y,\Tilde{y} \in E$ and $s,t \in I$ with $s<t$ we have:

\begin{align}\label{fundamental-calculus}
    \begin{split}
    \abs[\Big]{e^{-\frac{(x_t - \Tilde{y}_s)^2}{2C_\kappa(s,t)}} - e^{-\frac{(x_t - y_s)^2}{2C_\kappa(s,t)}}} 
     &= \frac{1}{C_\kappa(s,t)}\abs[\Big] 
     {\int_{y_s}^{\Tilde{y}_s}\textnormal{du}(u-x_t)e^{-\frac{(u-x_t)^2}{2C_\kappa(s,t)}}} \\
     &\leq
     \frac{M}{\sqrt{C_\kappa(s,t)}}|y_s - \Tilde{y}_s|.
    \end{split}
\end{align}

The equality is just the fundamental theorem of calculus applied to the function $y \mapsto e^{-\frac{(x_t - y)^2}{2C_\kappa(s,t)}}$ between the points $y_s$ and $\Tilde{y}_s$. As to the inequality, it follows from the basic fact that $xe^{-x^2}\leq M \equiv \frac{e^{-\frac{1}{4}}}{2}$ for all $x \geq 0$.

This leads to

\begin{align}
\begin{split}
    \abs[\Big]{
    F_\epsilon(x,y)(t) - F_\epsilon(x,\Tilde{y})(t)
    }
    &\lesssim 
    \int_0^{t-\epsilon}\textnormal{ds}
    \frac{m_s e^{\kappa s}}{C_\kappa(s,t)} |y_s -\Tilde{y}_s| 
    \\
    &\lesssim \frac{1}{\epsilon} \lVert m \rVert _\infty 
    \int_0^{t-\epsilon}\textnormal{ds}|y_s - \Tilde{y}_s|.
\end{split}
\end{align}

The first line is the triangle inequality, and the second is because 
$s \mapsto \frac{e^{\kappa s}}{C_\kappa (s,t)}$ is bounded above by $\frac{1}{\epsilon}$ on the interval $[0,t-\epsilon]$. This is the key argument that allows the proof to hold over $[0,t-\epsilon]$, but does not hold on $[0,t]$, and necessitated the introduction of our $\epsilon$-restrictions.

It follows that the iterated compositions of $F_\epsilon$ satisfy
\begin{align*}
    \abs[\Big]{F_\epsilon^n(x,y)(t) - F_\epsilon^n (x,\Tilde{y}(t)}
    \leq 
    \frac{C}{\epsilon} \int_0 ^{t-\epsilon}\textnormal{ds}
    \abs[\Big]{F_\epsilon^{n-1}(x,y)(t) - F_\epsilon^{n-1} (x,\Tilde{y}(t)}
\end{align*}

for some finite real constant, which in turn entails that

\begin{align*}
    \lVert F_\epsilon^n(x,y) - F_\epsilon^n (x,\Tilde{y}) \rVert_\infty
    \leq
    \frac{1}{n!}\left( \frac{C(T-\epsilon)}{\epsilon} \right)^n \lVert y - \Tilde{y}\rVert_\infty.
\end{align*}

On the right-hand side we find the general term of an exponential series, which must be less than $1$ for sufficiently large $n$. Then, for any such value of $n$, $F_\epsilon^n(x,\cdot)$ is a contraction in the Banach space E. This guarantees that $F_\epsilon(x,\cdot)$ has a unique fixed point $y_\epsilon(x)\in E$.

Now observe that the family $\left(y_\epsilon(x)\right)_{0 < \epsilon < T}$ is uniformly bounded with respect to the $L_\infty$-norm. It follows that for any $t \in I$, there exists a sequence $\epsilon_n \to 0$ such that $y_{\epsilon_n}(x)(t)$ converges as $n \to \infty$. Denoting its pointwise limit $y_0(x)(t)$, we obtain a measurable and bounded function $y_0(x)$.

Since for all $t \in I$,

\begin{align*}
\begin{split}
    \abs[\Big]{
    y_\epsilon(x)(t) - F_0(x,y_\epsilon(x))(t)}
    &= \abs[\Big]{ 
    F_\epsilon(x,y_\epsilon(x))(t) - F_0(x,y_\epsilon(x))(t)} \\
    &= \abs[\Big]{ 
    \int _{t-\epsilon}^t \frac{\textnormal{ds}m_s}{\sqrt{C_\kappa(s,t)}}e^{-\frac{
    \left( x-y_\epsilon(x)(s) \right)^2}{2C_\kappa(s,t)}}} \\
    &\lesssim 2 \lVert m \rVert _\infty \sqrt{\epsilon},
\end{split}
\end{align*}

we obtain by letting $\epsilon \to 0$ that $y_0(x)$ is a fixed point of $F_0(x, \cdot)$. One may observe that $y_0$ is continuous since it is the integral of a measurable function (and in fact is even infinitely differentiable).
This concludes the first part of the proof.

The second idea is to fix a time step $\delta$ and, starting from any function $x_0 \in \mathcal{C}([0,\delta],\mathbb{R})$, to iterate on the above fixed point procedure so as to build a solution to equation \eqref{impacted-price} of interest. Starting from $x_0$, one builds the fixed point $x_1 = y(x_0)$. One then iterates this procedure $n = \floor{\frac{T}{\delta}}$ times, giving a sequence of functions $(x_i)_{0\leq i \leq n}$. Finally, one concatenate these functions together; or, more explicitly, consider the function $y_\delta$ defined on $[0,T]$ by $y_\delta(t) = x_k(t-k\delta)$ where $k = \floor{\frac{t}{\delta}}$.

By construction, the function $y_\delta$ satisfies

\begin{align*}
    y_\delta(t) = \int_0^t \frac{\textnormal{ds}m_s}{\sqrt{2\pi C_\kappa(s,t)}}
    e^{-\frac{\left(y_\delta(t-\delta) - y_s \right)^2 }{2C_\kappa(s,t)}}.
\end{align*}

We obtain the desired solution by letting $\delta \to 0$.
\end{proof}

\end{document}